\documentclass[a4paper,11pt]{article}

\usepackage{jinstpub} 
\usepackage{graphicx}
\usepackage{subfigure}
\usepackage{verbatim}
\usepackage{url}

\title{\boldmath Development of a high-granularity, high-precision timing readout electronics system for large-area MRPC detectors}

\author[a,b]{J.N. Tang,}
\author[a,b,1]{W.H. Wu,\note{Corresponding author.}}
\author[a,b]{Y.Q. Tan,}
\author[a,b]{W. Zhi,}
\author[a,b]{H.J. Yang,}
\author[c]{Imad Laktineh,}
\author[d]{Q.P. Shen}

\affiliation[a]{Institute of Nuclear and Particle Physics, School of Physics and Astronomy, Shanghai Jiao Tong University,\\800 Dongchuan Road, Shanghai, China}
\affiliation[b]{Key Laboratory for Particle Astrophysics and Cosmology (MoE), Shanghai Key Laboratory for Particle Physics and Cosmology,\\800 Dongchuan Road, Shanghai, China}
\affiliation[c]{Institut de Physique des 2 Infinis de Lyon, Lyon University\\69622 Villeurbanne, Lyon, France}
\affiliation[d]{Institute of High Energy Physics, Chinese Academy of Sciences,\\19B Yuquan Road, Beijing, China}

\emailAdd{wuweihao@sjtu.edu.cn}

\abstract{Multi-gap Resistive Plate Chamber (MRPC) detectors offer excellent time resolution and detection efficiency, creating a strong demand for high-precision, highly scalable timing readout systems. In this work, a readout electronics system is designed for a large-area ($100 \times 100\text{~cm}^2$) MRPC detector containing 2400 high-granularity $2 \times 2\text{~cm}^2$ pad channels. The system consists of two Front-End Boards (FEBs), a central clock distribution module, and a back-end DAQ aggregator. Each FEB is equipped with 40 32-channel PETIROC2B ASICs mounted directly behind the sensing pads. An automated S-curve calibration procedure equalizes the baseline dispersion across all 2400 channels, reducing the FWHM of the baseline voltage distribution from 50~mV to 12~mV and establishing a uniform triggering threshold. Signal-injection measurements confirm an intrinsic single-channel electronic time resolution of $\sigma_{\text{intra-chip}} \approx 33\text{~ps}$ RMS, alongside inter-chip and inter-board time resolutions of $\approx 43\text{~ps}$ RMS and $\approx 45\text{~ps}$ RMS, respectively. This high-granularity, high-precision timing readout system can be widely applied to Time-of-Flight systems, cosmic-ray muon imaging, as well as other fast-timing detector systems.}

\keywords{Front-end electronics for detector readout; Gaseous detectors; Modular electronics}

\arxivnumber{xxxx.xxxxx}

\begin{document}
\maketitle
\flushbottom

\section{Introduction}
\label{sec:intro}

Multi-gap Resistive Plate Chambers (MRPCs) evolved from traditional Resistive Plate Chambers (RPCs)~\cite{akindinov2000multigap}. By partitioning the gas volume into multiple sub-millimeter sub-gaps (typically $100\text{--}250\text{~}\mu\text{m}$) with floating resistive plates, MRPCs maintain high, uniform internal electric fields while drastically reducing the avalanche growth time. This architecture delivers superior time resolutions ($<100\text{~ps}$) and high detection efficiencies ($>95\%$)~\cite{wang2019status}. Consequently, MRPCs have become indispensable for Time-of-Flight (TOF) systems in large-scale high-energy physics experiments (such as STAR~\cite{llope2012multigap}, ALICE~\cite{akindinov2013performance}, and CBM~\cite{wang2016development}), while increasingly finding applications in medical imaging (e.g., TOF-PET~\cite{amaldi2015development}) and cosmic-ray muon tomography~\cite{wang2015cosmic}. Leveraging these excellent timing capabilities, MRPC detectors are also proposed to replace conventional RPCs in the high-granularity Semi-Digital Hadronic Calorimeter (SDHCAL) concept~\cite{beaulieu2015conception} for future Higgs factories, such as FCC-ee. The newly proposed Timing-SDHCAL (T-SDHCAL)~\cite{tytgat2025towards} exploits the precise arrival time of hits created by hadronic showers in the MRPC, enabling superior shower separation and reconstruction through Particle Flow Algorithm (PFA) techniques to ultimately improve jet energy resolution.

The choice of readout topology and front-end electronics determines the throughput and timing bounds of the MRPC:
\begin{itemize}
    \item \textbf{Strip vs.\ Pad Readout:} Double-ended strip readouts minimize channel counts by extracting hit positions from differential arrival times~\cite{sun2008new}. However, at high particle fluxes, they suffer from hit ambiguity and severe pileup. High-granularity pad arrays eliminate spatial confusion and isolate high-rate events~\cite{wang2013development}, though at the expense of significantly higher channel density.
    \item \textbf{Electronics Architectures:} Switched Capacitor Array (SCA) digitization offers complete waveform profiling but is limited by high power dissipation, massive data volume, and restricted channel density~\cite{wang2012design}. In contrast, the front-end ASIC plus TDC/TAC paradigm integrates preamplifiers and fast discriminators~\cite{anghinolfi2003nino}, delivering superior channel scalability, low power consumption, and direct timing measurement ideal for dense pad readout.
\end{itemize}

For the large-area MRPC ($100 \times 100\text{~cm}^2$) developed in this study, a high-granularity $2 \times 2\text{~cm}^2$ pad matrix was chosen to meet demanding spatial resolution and high-rate requirements, yielding a total of 2400 independent readout channels. Routing 2400 high-frequency, small-amplitude analog signals (on the order of femtocoulombs) via external coaxial cables or dense connectors introduces severe signal attenuation, parasitic capacitance, crosstalk, and mechanical integration unfeasibility. 

To overcome this bottleneck, an active front-end PCB architecture was designed: the front side features the $2 \times 2\text{~cm}^2$ pad matrix, while the reverse side hosts the integrated readout electronics directly behind the pads, minimizing stray capacitance and signal path lengths. To accommodate 2400 channels under strict space and thermal constraints, the front-end ASIC must provide high channel density, low power dissipation, and integrated on-chip digitization for both time and charge (to enable offline time-slewing correction). Furthermore, onboard FPGAs are required for local data aggregation, slow control, and high-speed serialized data transmission.

In this paper, we present the design and implementation of an ASIC-based integrated readout electronics system tailored for a $100 \times 100\text{~cm}^2$ MRPC detector. Section~\ref{sec:architecture} details the system architecture, ASIC selection rationale, and system-level synchronization topology.


\section{Architecture of Readout Electronics System}
\label{sec:architecture}

The readout system is structured to manage 2400 high-density pad channels while ensuring sub-50~ps timing alignment across the entire $100 \times 100\text{~cm}^2$ detector surface.

\subsection{PETIROC2B ASIC Introduction}
\label{sec:asic}

The PETIROC2B ASIC~\cite{fleury2013petiroc,fleury2014petiroc} was selected as the core front-end chip due to its high integration, low power consumption ($\sim 6\text{~mW/channel}$), and high-rate capability (up to $40\text{~kHz}$). Designed for 32-channel parallel processing, each analog input path inside the chip is split into two functional branches upon entry: a fast timing branch and a charge measurement branch. 

In the timing path, the signal passes through a high-bandwidth preamplifier and a fast discriminator to trigger an integrated Time-to-Amplitude Converter (TAC), delivering an intrinsic time resolution of approximately $37\text{~ps}$ RMS. To optimize triggering performance and correct channel-to-channel baseline variations, the timing discriminator threshold for each channel is precisely defined by combining a 10-bit global DAC ($V_{\text{th\_time}}$) with an individual 6-bit channel-wise Trim-DAC. 

In parallel, the charge measurement path processes the shaped signal to enable offline time-slewing correction over a wide dynamic range. Both digitized time and charge information are converted on-chip, eliminating the need for external discrete TDCs or ADCs, thereby significantly relaxing PCB routing complexity and power density constraints across the large-area detector.

\subsection{Modular System Topology}
\label{sec:topology}

\begin{figure}[htbp]
\centering
\includegraphics[width=0.98\textwidth]{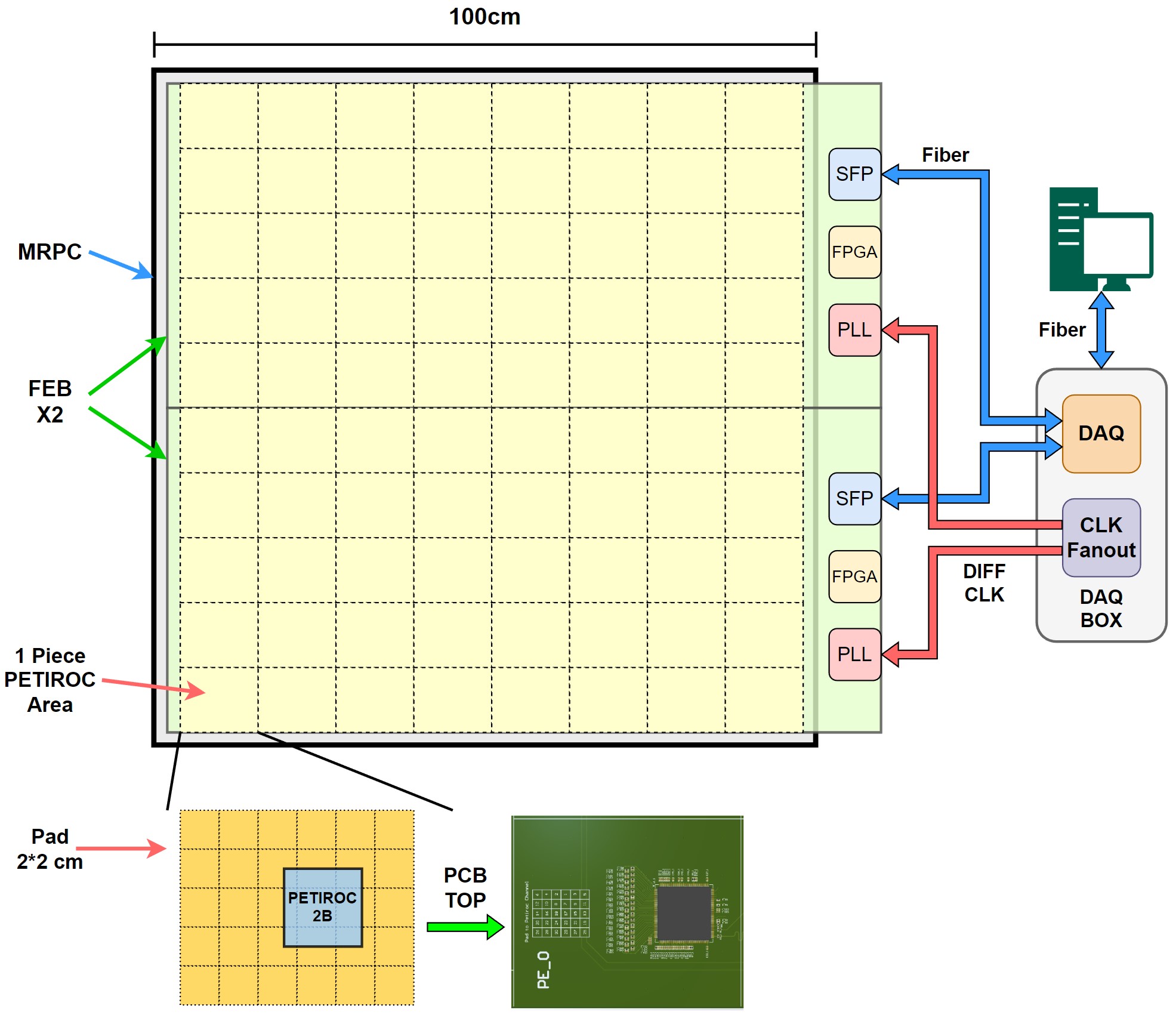}
\caption{Block diagram of the MRPC readout electronics system: two FEBs (indicated by green arrows) are mounted directly on top of the MRPC detector (indicated by blue arrow). The yellow dashed box (indicated by red arrow) shows the readout area of a single PETIROC2B. The two FEBs exchange data with the back-end DAQ module via optical links (SFP), while a clock fanout module delivers synchronized clocks to both FEBs.}
\label{fig:sys_layout_block}
\end{figure}

Due to manufacturing constraints for large-scale, multi-layer PCBs, fabricating a single $100 \times 100\text{~cm}^2$ active board is cost-prohibitive. Consequently, the active readout area is divided into two identical FEBs, each covering a $100 \times 50\text{~cm}^2$ region and serving 1200 pads. As illustrated in figure~\ref{fig:sys_layout_block}, two FEBs are mounted directly on top of the MRPC detector.

The readout electronics consist of three hardware modules: two FEBs, a central clock distribution module, and a DAQ module. Each FEB reads out 1200 pads arranged in a high-density matrix. On the reverse side, 40 PETIROC2B ASICs are placed adjacent to their corresponding pad groups to read out the MRPC signal. A single Xilinx Kintex-7 FPGA (XC7K325T) on each FEB serves as the local controller, managing ASIC configuration and data reception.

To achieve global time alignment across all 2400 channels, a clock distribution module delivers low-jitter clock and synchronous signals to both FEBs via phase-matched, shielded differential lines. This centralized topology suppresses inter-board timing skew and maintains phase stability. Each FEB transmits aggregated event packets to the back-end DAQ Aggregator Module via a 10~Gbps optical link (SFP+). The DAQ module merges data streams from both FEBs and relays data to the host PC via 10-Gigabit Ethernet using the SiTCP~\cite{uchida2007hardware} protocol.


\section{Front-End Board Hardware Design}
\label{sec:feb_design}

\subsection{Global FEB Architecture and Physical Layout}
\label{sec:feb_layout}

\begin{figure}[htbp]
\centering
\includegraphics[width=0.95\textwidth]{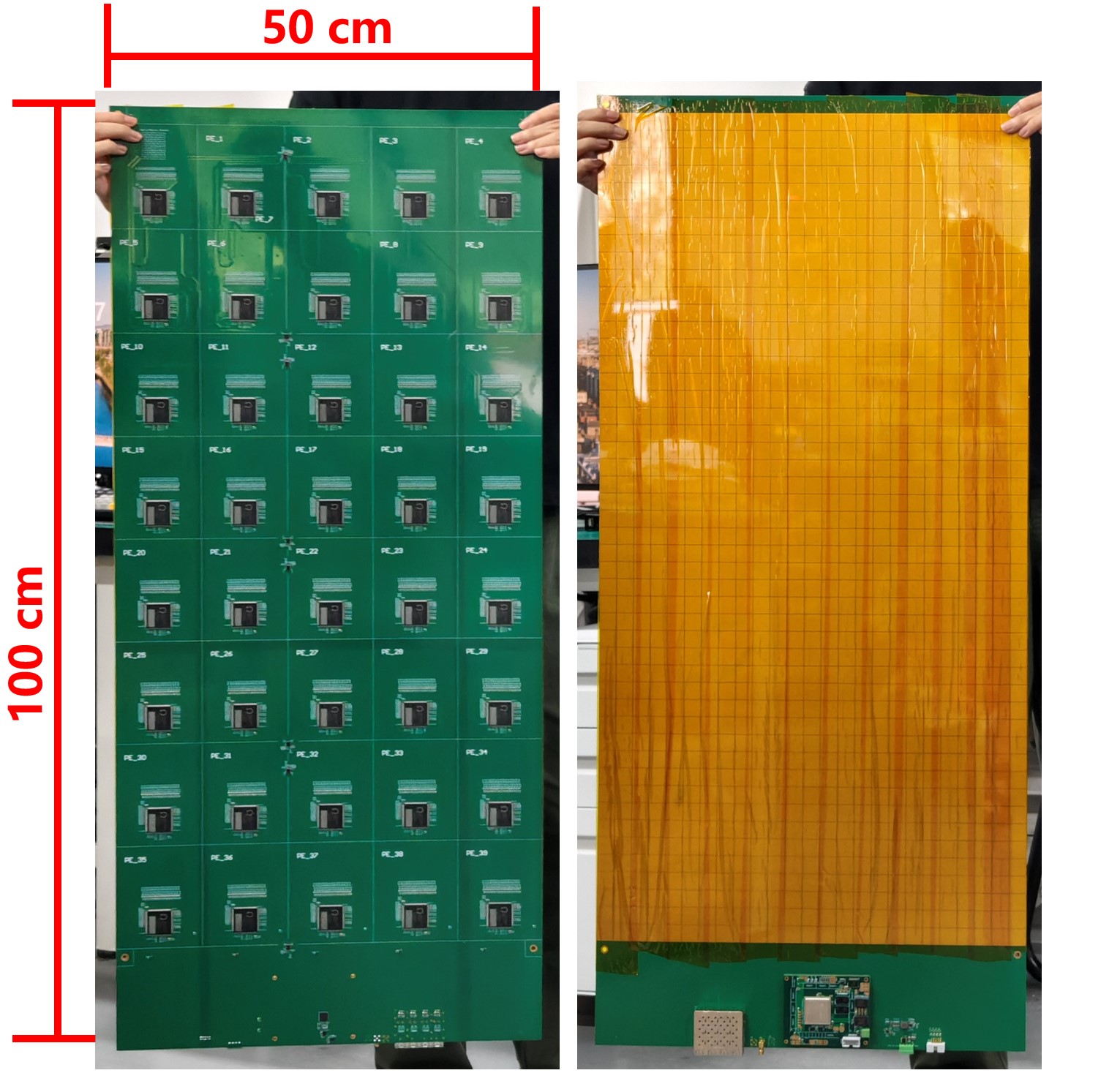}
\caption{Photographs of the Front-End Board (FEB): top side (left) and bottom side (right). The top side houses 40 PETIROC2B ASICs. The bottom side features the 2400-channel readout pad matrix directly beneath the ASICs, with the FPGA core board and optical connectors (SFP) located in the lower section.}
\label{fig:PCB_Photo_TopAndBot}
\end{figure}

To read out the active $100 \times 100\text{~cm}^2$ MRPC detector, two identical Front-End Boards (FEBs), each measuring $110 \times 50\text{~cm}^2$, are utilized side by side. Each FEB handles a $96 \times 50\text{~cm}^2$ detection area accommodating 1200 readout pads ($2 \times 2\text{~cm}^2$ each), arranged in a 25-column by 48-row matrix. Figure~\ref{fig:PCB_Photo_TopAndBot} shows the physical layout of the ASICs, readout pads, and main components on the FEB.

Physically, each FEB is logically partitioned into two functionally distinct regions to minimize crosstalk: \textbf{Active Sensing Region} ($100 \times 50\text{~cm}^2$) and \textbf{Control and Processing Region} ($10 \times 50\text{~cm}^2$). The Active Sensing Region is located on the main detector coverage area. The front layer accommodates the 1200 readout pads, while the reverse layer hosts a $5 \times 8$ array of PETIROC2B ASICs (40 chips in total), as shown in figure~\ref{fig:PCB_Photo_TopAndBot}. Each PETIROC2B directly interfaces with a localized $5 \times 6$ pad subarray ($10 \times 12\text{~cm}^2$), establishing a compact, direct-coupling signal pathway. The bottom of figure~\ref{fig:sys_layout_block} shows the readout area of one PETIROC2B.

The Control and Processing Region extends beyond the active MRPC boundaries to avoid physical mechanical interference and sensitive analog coupling. This region integrates a custom Xilinx Kintex-7 (XC7K325T) FPGA core board attached via high-speed board-to-board connectors, a low-jitter clock management subsystem, a low-noise Power Network, and an SFP+ optical transceiver module connected directly to the FPGA GTX transceivers, as shown in figure~\ref{fig:PCB_Photo_TopAndBot}.

\subsection{High-Signal-Integrity PCB Design}
\label{sec:pcb_design}

\begin{figure}[htbp]
	\centering
	\subfigure[]{
  	    \includegraphics[width=.455\textwidth]{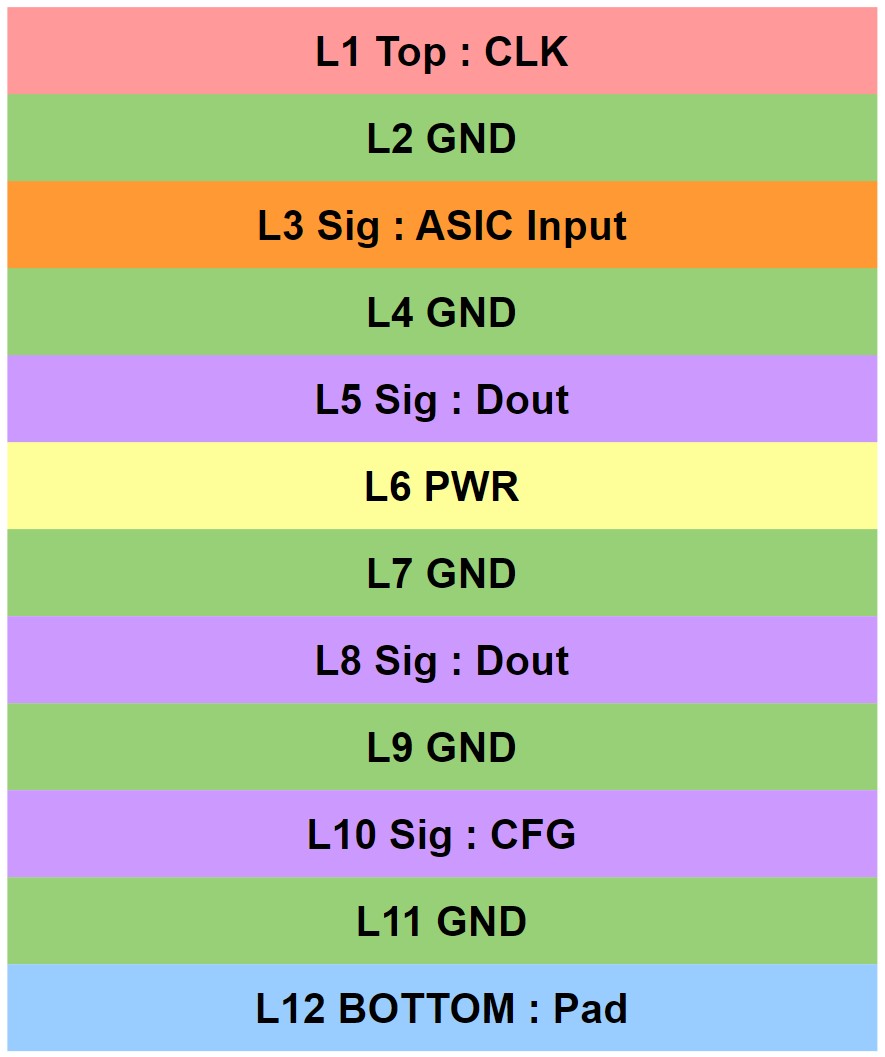}
   	    \label{fig:feb_layer_setup}
    }
    \subfigure[]{
  	    \includegraphics[width=.45\textwidth]{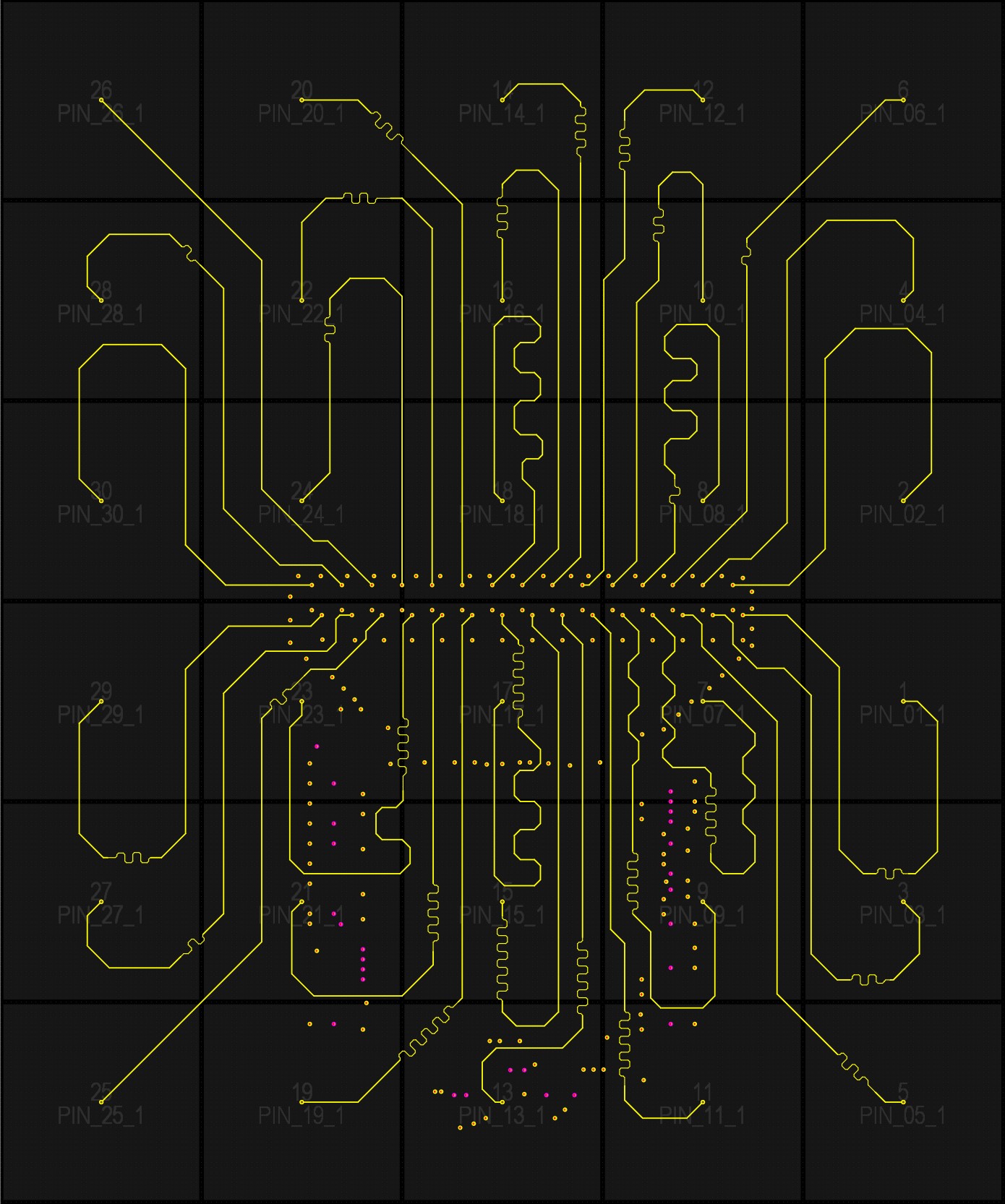}
   	    \label{fig:ana_input_trace}
    }
    \caption{\label{fig:feb_pwr_clk_block}Figure~\ref{fig:feb_layer_setup} shows the layer-stackup of the FEB, while figure~\ref{fig:ana_input_trace} illustrates the 30 analog (yellow) traces from readout pads in Layer 3.}
\end{figure}

Because the MRPC induced charges are extremely small (on the order of femtocoulombs) and the PETIROC2B analog inputs are sensitive to electromagnetic interference, preserving signal integrity is paramount. The PCB design isolates fragile analog signals from digital switching noise through physical segregation and rigorous layer-stackup shielding, as shown in figure~\ref{fig:feb_layer_setup}.

Within each $10 \times 12\text{~cm}^2$ functional unit, signals from the readout pads on the bottom layer transition to Layer 3 via blind/buried micro-vias placed directly at the pad centers. Layer 3 is strictly reserved for routing analog signals from pads to the corresponding ASIC input pins. This layer forms a stripline structure shielded above and below by solid ground (GND) planes on Layer 2 and Layer 4. Furthermore, as shown in figure~\ref{fig:ana_input_trace}, all 30 analog traces within each unit are length-matched to ensure identical signal propagation delay across all channels, suppressing inter-channel timing skew. Digital control lines and high-speed serialized data streams are routed strictly on inner layers completely encased by continuous GND planes, providing $360^\circ$ shielding against crosstalk into the analog front-end.

\subsection{ASIC Configuration and Readout Control}
\label{sec:asic_control}

PETIROC2B contains an internal 640-bit shift register for channel configuration and threshold setting, accessed via a serial input (\texttt{SR\_IN}) and output (\texttt{SR\_OUT}). To balance FPGA I/O utilization and timing closure constraints, the 40 ASICs are divided into 5 columns, each containing 8 cascaded ASICs in a daisy-chain topology. The \texttt{SR\_OUT} of the final ASIC in each chain is looped back to the FPGA. Configuration data is shifted through the chain and read back iteratively, allowing the FPGA firmware to perform real-time bitwise verification to ensure correct register configuration.

During operational acquisition, an internal OR-logic combines the outputs of all 32 channel discriminators within each ASIC. Upon detecting a hit, the fast trigger signal alerts the FPGA. The FPGA subsequently issues a \texttt{start\_conv} pulse to initiate on-chip time and charge digitization. Once conversion finishes, each channel outputs a 30-bit frame (960 bits per ASIC). The serialized data stream is transmitted to the FPGA via differential \texttt{dout} lines at 80~Mbps, synchronized by a frame-valid signal (\texttt{transmit\_on}). All 40 \texttt{dout} lines are connected directly to dedicated LVDS pairs on the Kintex-7 FPGA for parallel data capture.

\subsection{Low-Jitter Clock Distribution System}
\label{sec:clock_tree}

To maintain sub-50~ps timing precision across the 1200 channels of an FEB, all 40 ASICs require identical, low-jitter, and phase-aligned clocks.

\begin{figure}[htbp]
\centering
\includegraphics[width=0.80\textwidth]{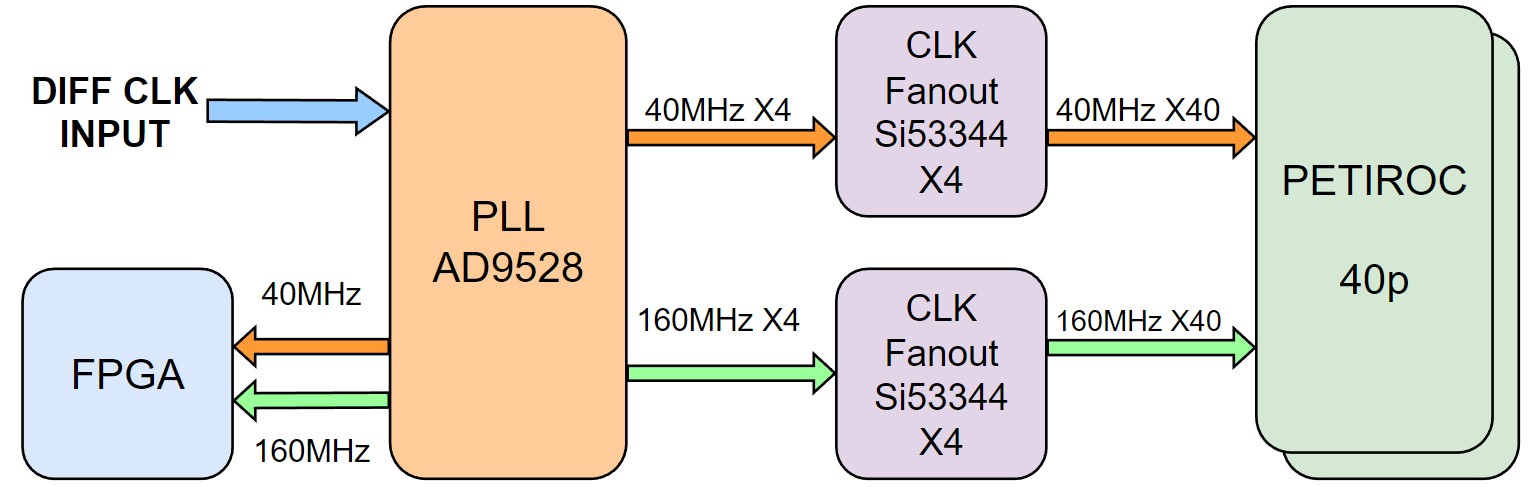}
\caption{Block diagram of the FEB clock distribution system.}
\label{fig:feb_clk_block}
\end{figure}

The clock distribution system employs a two-stage distribution architecture, as shown in figure~\ref{fig:feb_clk_block}. An external differential reference clock is received by an onboard jitter-cleaner PLL (AD9528), suppressing clock jitter below 200~fs RMS. The AD9528 generates 5 phase-aligned differential clock pairs at 40~MHz and 160~MHz. One pair is routed to the FPGA for coarse timestamping and data deserialization, while the remaining 4 pairs drive four 1-to-10 clock fanout buffers (Si53344). These buffers generate 40 synchronous 40~MHz and 160~MHz differential clock pairs distributed to each PETIROC2B. The Si53344 exhibits an ultra-low output-to-output skew ($<20\text{~ps}$) and minimal additive jitter ($<50\text{~fs}$). Oscilloscope measurements confirm phase synchronization across all 40~MHz ASIC clock lines, satisfying strict timing distribution requirements.

\subsection{Low-Noise Power Network}
\label{sec:power_design}

\begin{figure}[htbp]
\centering
\includegraphics[width=0.70\textwidth]{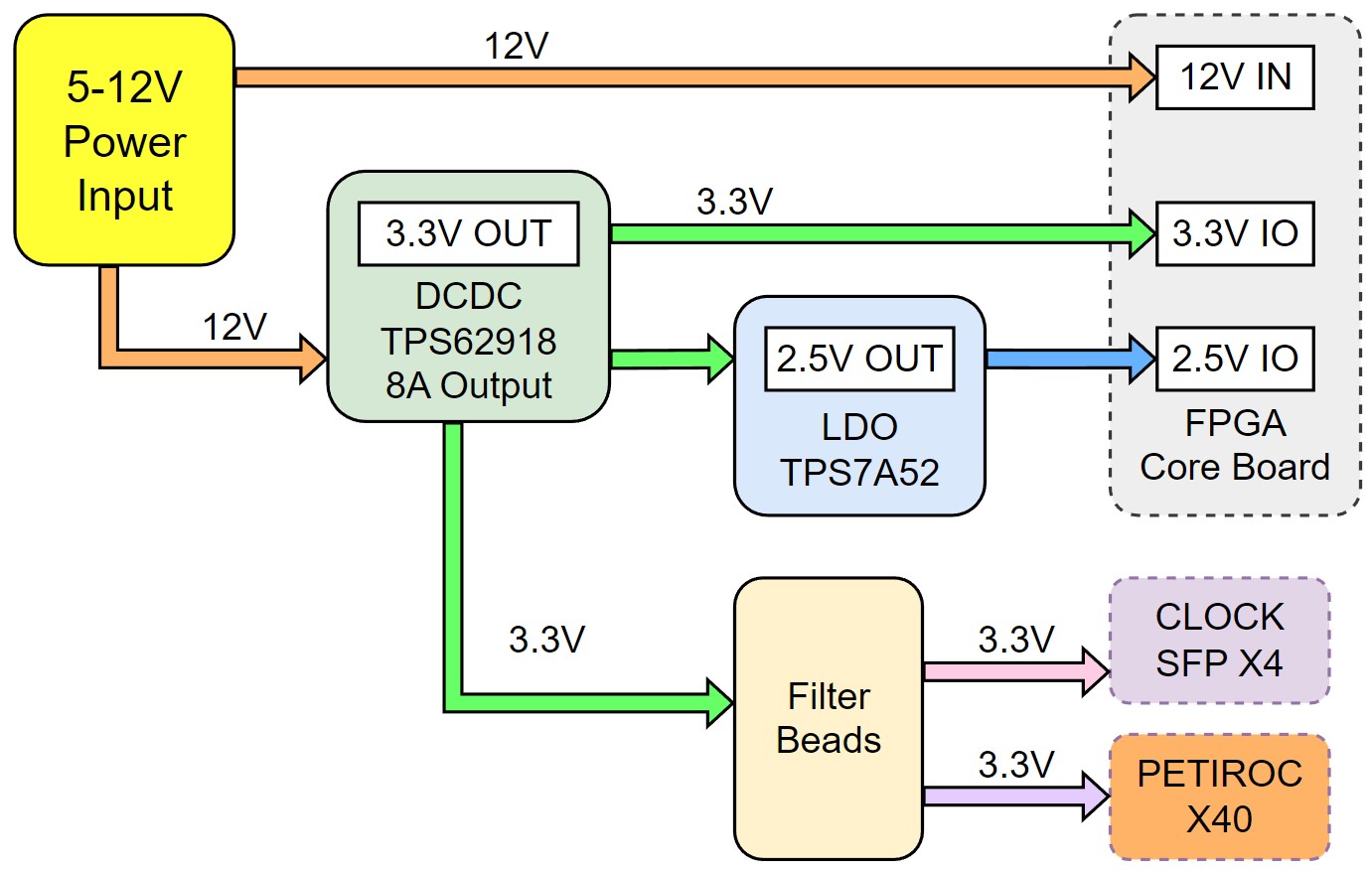}
\caption{Block diagram of the FEB power network.}
\label{fig:feb_pwr_block}
\end{figure}

The power network of the FEB is shown in Figure~\ref{fig:feb_pwr_block}. The low-noise preamplifiers and ultra-fast discriminators of the PETIROC2B, as well as the AD9528 clock PLL, demand a highly stable power supply. Operating from a 5--12~V main supply input, the FEB requires substantial 3.3~V supply current for the 40 ASICs and clock circuitry. Traditional LDO regulators would cause excessive thermal dissipation due to low power efficiency.

To address this, a high-efficiency buck switching regulator (TPS62918) with low output ripple ($<1\text{~mV}$) is selected to step down the input voltage to 3.3~V. To prevent switching noise propagation into sensitive analog nodes, the 3.3~V output is split via magnetic beads into two isolated power domains: a dedicated low-noise 3.3~V rail powering the PETIROC2B ASICs and an isolated 3.3~V rail for peripheral digital circuitry. A dedicated internal power plane enclosed by adjacent ground planes is allocated for the ASIC 3.3~V supply, providing low impedance and superior noise decoupling.


\section{Firmware and Software Architecture}
\label{sec:firmware_software}

\begin{figure}[htbp]
\centering
\includegraphics[width=0.99\textwidth]{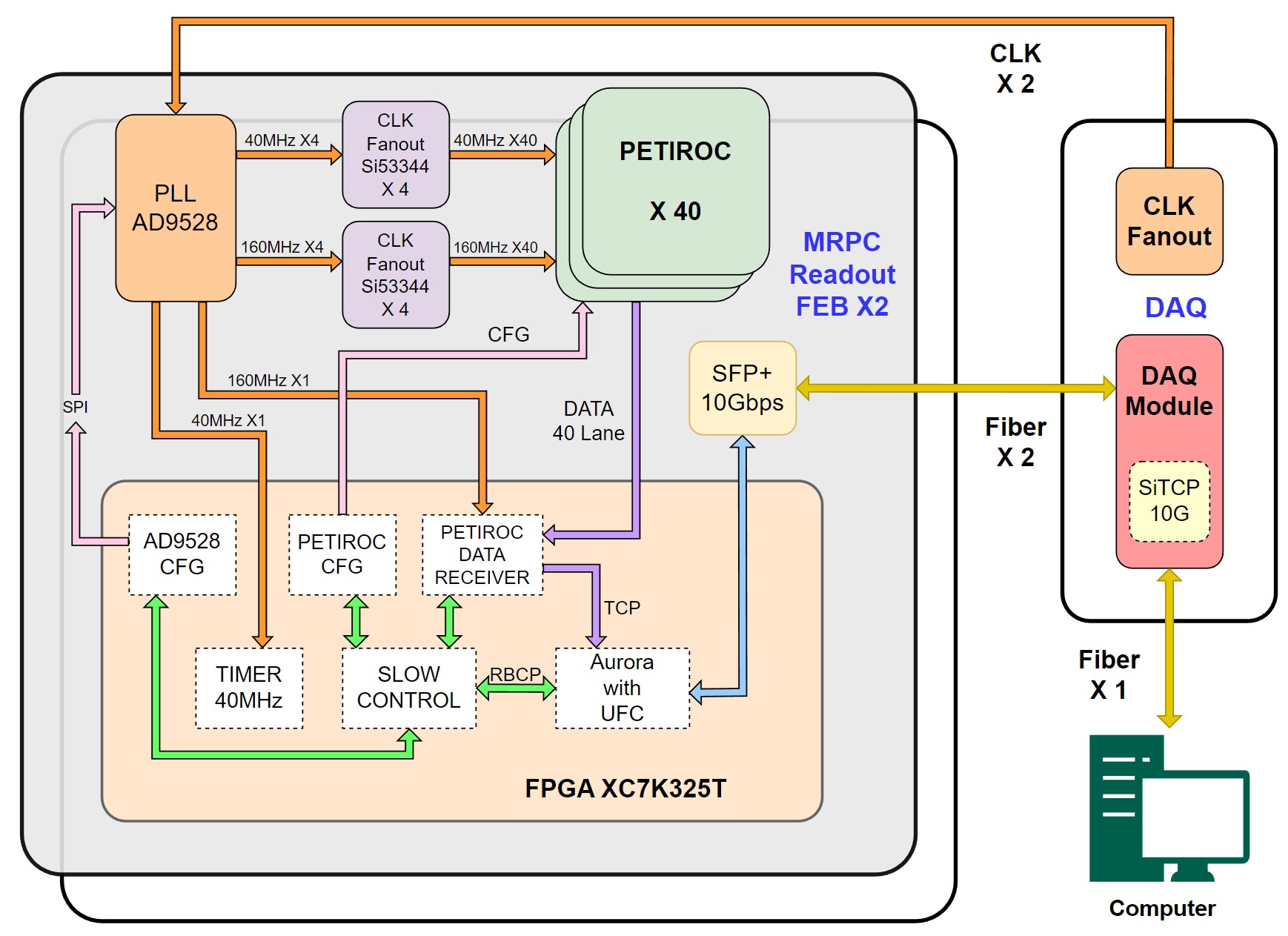}
\caption{Block diagram of the FEB firmware and DAQ system.}
\label{fig:sys_block}
\end{figure}

This section details the DAQ system architecture, which spans the onboard FPGA firmware on each FEB, the back-end DAQ concentrator module, and the host online software. The firmware architecture of the readout electronics is shown in Figure~\ref{fig:sys_block}.

\subsection{Front-End Data Reception and DAQ Firmware}
\label{sec:firmware_daq}

The data readout firmware is built as a multi-stage pipeline across the FEB and DAQ FPGAs to process random detector hits efficiently. It manages parallel serial data recovery, timestamp framing, multi-channel FIFO arbitration, and high-speed optical streaming.

Each PETIROC2B ASIC outputs serialized 960-bit event frames over a differential \texttt{dout} line at 80~Mbps upon asserting its frame-valid signal (\texttt{transmit\_on}). To overcome the lack of an accompanying synchronous data clock and compensate for trace propagation variations across the large $110 \times 50\text{~cm}^2$ FEB, strict length-matching is applied to all differential \texttt{dout} and \texttt{transmit\_on} traces, ensuring minimal relative timing skew at the FPGA boundary. An internal MMCM multiplies the 160~MHz reference clock from the AD9528 up to 320~MHz, providing a $4\times$ over-sampling clock relative to the 80~Mbps bit rate. Driven by the rising edge of \texttt{transmit\_on}, a 320~MHz sampling counter latches each bit at the maximum eye-diagram opening to maximize sampling noise margins, and the recovered bitstream is subsequently converted into 64-bit parallel words, yielding 15 words per 960-bit ASIC frame.

The PETIROC2B internal 9-bit coarse timestamp covers only $12.8~\mu\text{s}$. To prevent counter overflow during long runs, the FPGA maintains a global 56-bit coarse counter driven by a 40~MHz clock. For each 960-bit payload, the FPGA adds a 128-bit metadata header containing: a 48-bit sync pattern for frame alignment, the 16-bit DAC threshold ($V_{\text{th\_time}}$), the 56-bit extended timestamp, and an 8-bit ID (2-bit FEB ID and 6-bit ASIC ID) to identify the pad location. A CRC field is added to the end of the frame for data integrity checks.

To manage random detector hits, each of the 40 ASICs feeds an independent asynchronous FIFO. Each FIFO uses a programmable full threshold (\texttt{prog\_full}) to handle backpressure, leaving enough space to store a complete 960-bit frame and avoiding frame truncation. A round-robin scheduler monitors the 40 FIFOs and reads out complete frames whenever data is ready, merging them into a single concentrator FIFO connected to the Aurora optical interface.

\begin{figure}[htbp]
\centering
\includegraphics[width=0.75\textwidth]{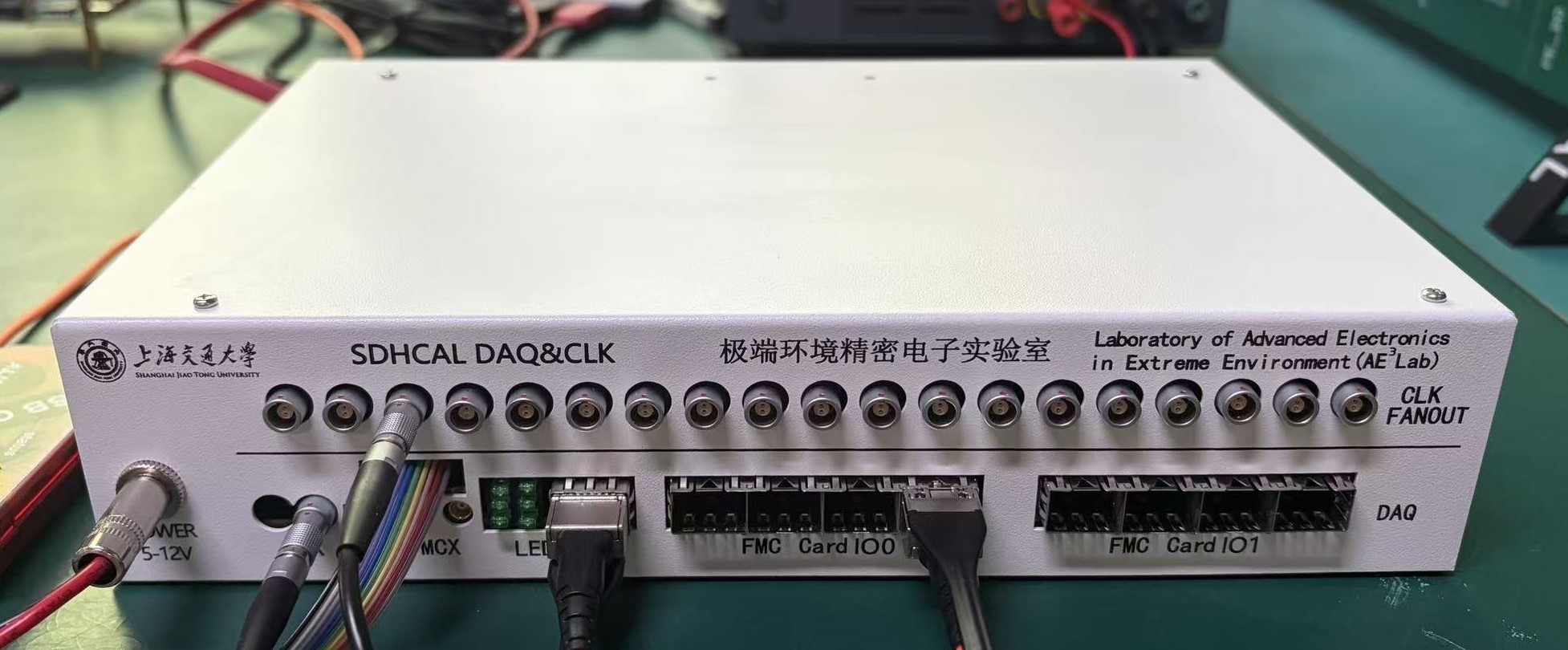}
\caption{Photographs of the back-end DAQ (bottom) and clock fanout (top) module.}
\label{fig:daq_module}
\end{figure}

The back-end DAQ module receives Aurora data streams from two FEBs ($100 \times 100\text{~cm}^2$ total detector area), as shown in Figure~\ref{fig:daq_module}. The DAQ checks the frame header and CRC checksum before saving valid data into primary FIFOs. A secondary round-robin scheduler reads full frames into a main concentrator FIFO. This FIFO connects directly to a 10-Gigabit SiTCP IP core, which streams the event data over a 10Gbps optical Ethernet link to the host computer using standard TCP/IP.

\subsection{Distributed Slow Control via SiTCP and Aurora UFC}
\label{sec:firmware_slow_control}

The slow control system manages system configuration and hardware status monitoring. It operates alongside the high-rate DAQ path without consuming physics data bandwidth by combining SiTCP RBCP and the Aurora User Flow Control (UFC)~\cite{amd2024aurora}.

The host computer sends control commands using UDP-based RBCP packets, which provide automatic acknowledgment and retry. Each RBCP packet carries a read/write command, a 32-bit address, and an 8-bit data byte. The DAQ module uses the top 8 bits of the address (\texttt{Address[31:24]}) as a board ID to route the command to the target FEB. Upon receiving the command, the DAQ module combines the read/write bit, local address, and 8-bit data into a simple 64-bit control frame and transmits it over Aurora UFC. Because Aurora UFC has a higher priority than normal physics data streaming, it temporarily pauses physics data transfer to send the control frame immediately with low latency, and then resumes data transfer.

On the FEB, the incoming 64-bit UFC frame directly accesses a local dual-port RAM that acts as a register map. This RAM manages writable configuration parameters—including the 640-bit shift registers for the 40 PETIROC2B ASICs (organized in 5 daisy chains with 8 ASICs per chain), system reset logic, and DAC threshold settings ($V_{\text{th\_time}}$ and $V_{\text{th\_charge}}$). It also stores read-only status data, such as FPGA temperature (sampled by XADC), AD9528 PLL lock status, power rail health, and optical link quality. Once a command is executed, the FEB sends a response back to the DAQ through UFC, and the DAQ returns an RBCP ACK packet to the host computer to complete the control process.

\subsection{Online Control and Calibration Software}
\label{sec:online_software}

A GUI software built with PyQt and Python Socket networking handles data logging and system control. The software implements two main network interfaces: a TCP client socket connected to the SiTCP data port for high-throughput disk logging and real-time hit rate monitoring, and an RBCP UDP socket connected to the control port for parameter configuration.

To calibrate discriminator thresholds across all 2400 channels, an automated S-curve scanning module is embedded in the software. The application sequentially adjusts the 10-bit $V_{\text{th\_time}}$ DAC registers via RBCP, triggers internal test pulses or baseline noise measurements, and records channel hit counts as a function of threshold voltage. The resulting S-curves are fitted online using an Error Function (ERF) to extract channel baseline noise ($\sigma$) and determine the $50\%$ trigger threshold point ($V_{50\%}$), enabling automated threshold equalization across the entire active area.


\section{Performance Evaluation and System Integration Tests}
\label{sec:performance_tests}

To evaluate the operational stability and noise performance of the designed front-end electronics, a series of comprehensive tests were conducted. A multi-channel calibration procedure was first implemented to equalize discriminator thresholds across the 2400 active readout channels, followed by system timing performance of readout electronics.

\subsection{Front-End Threshold Calibration and Channel Equalization}
\label{sec:threshold_calibration}

Setting a low, uniform operational threshold across all 2400 channels is essential to maximize MRPC detection efficiency while avoiding false noise triggers. An automated calibration procedure was developed to measure baseline noise for each channel and adjust individual thresholds using internal Trim-DACs.

The host PyQt software automatically steps the 10-bit global discriminator DAC ($V_{\text{th\_time}}$) using RBCP commands to record the noise hit rate for each channel. Because adjacent $2 \times 2\text{~cm}^2$ pads can couple noise to each other during simultaneous switching, a row-by-row activation method is used. The software enables one row of pads at a time instead of all 32 channels on an ASIC. This spatial separation removes digital-to-analog crosstalk and ensures an accurate measurement of the true noise floor.

\begin{figure}[htbp]
\centering
\includegraphics[width=0.99\textwidth]{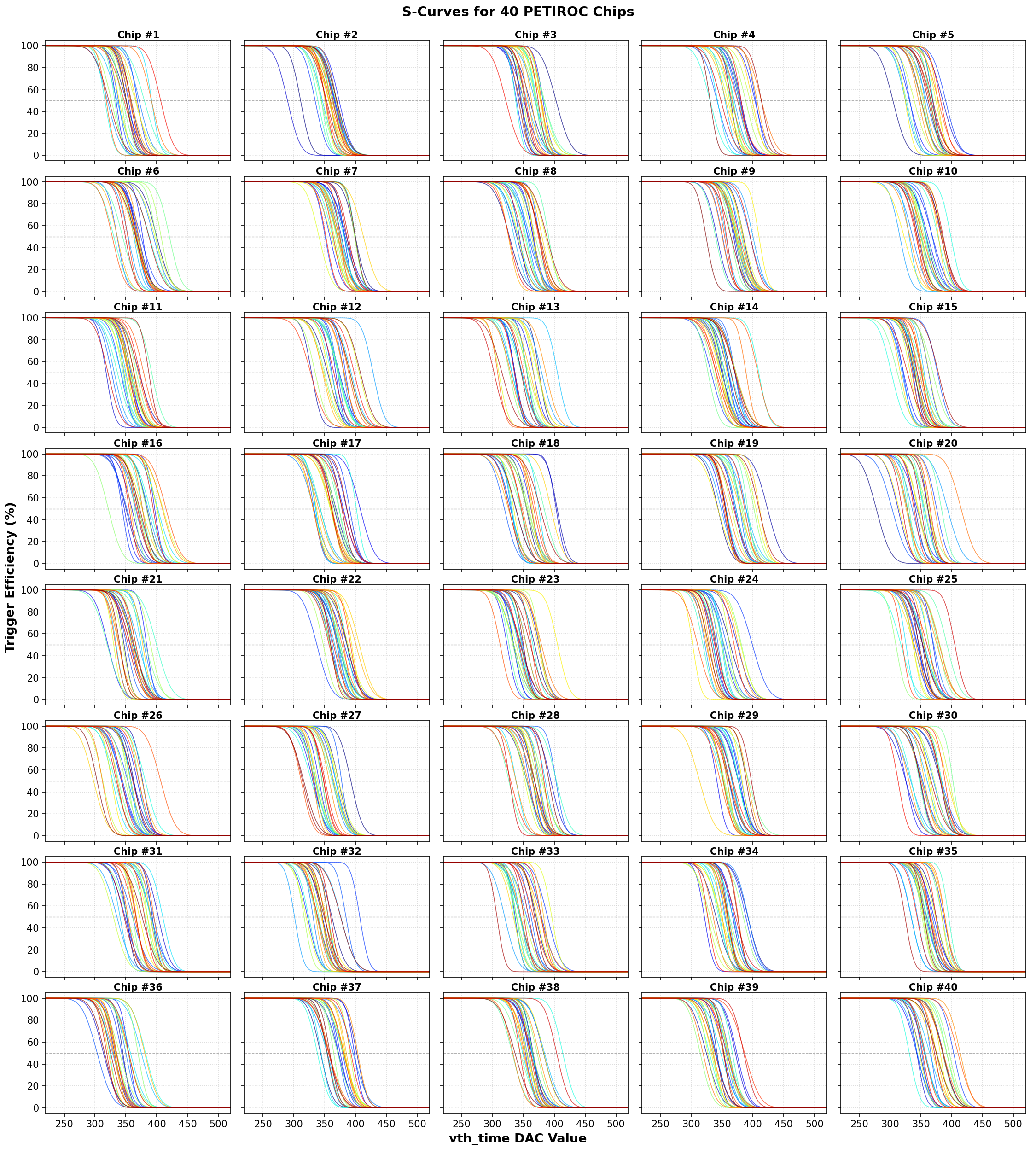}
\caption{The trigger efficiency of 40 PETIROCs as a function of threshold voltage $V_{\text{th}}$ before trim-DAC equalization.}
\label{fig:s_curve_before_cali}
\end{figure}

\begin{figure}[htbp]
	\centering
	\subfigure[]{
  	    \includegraphics[width=.65\textwidth]{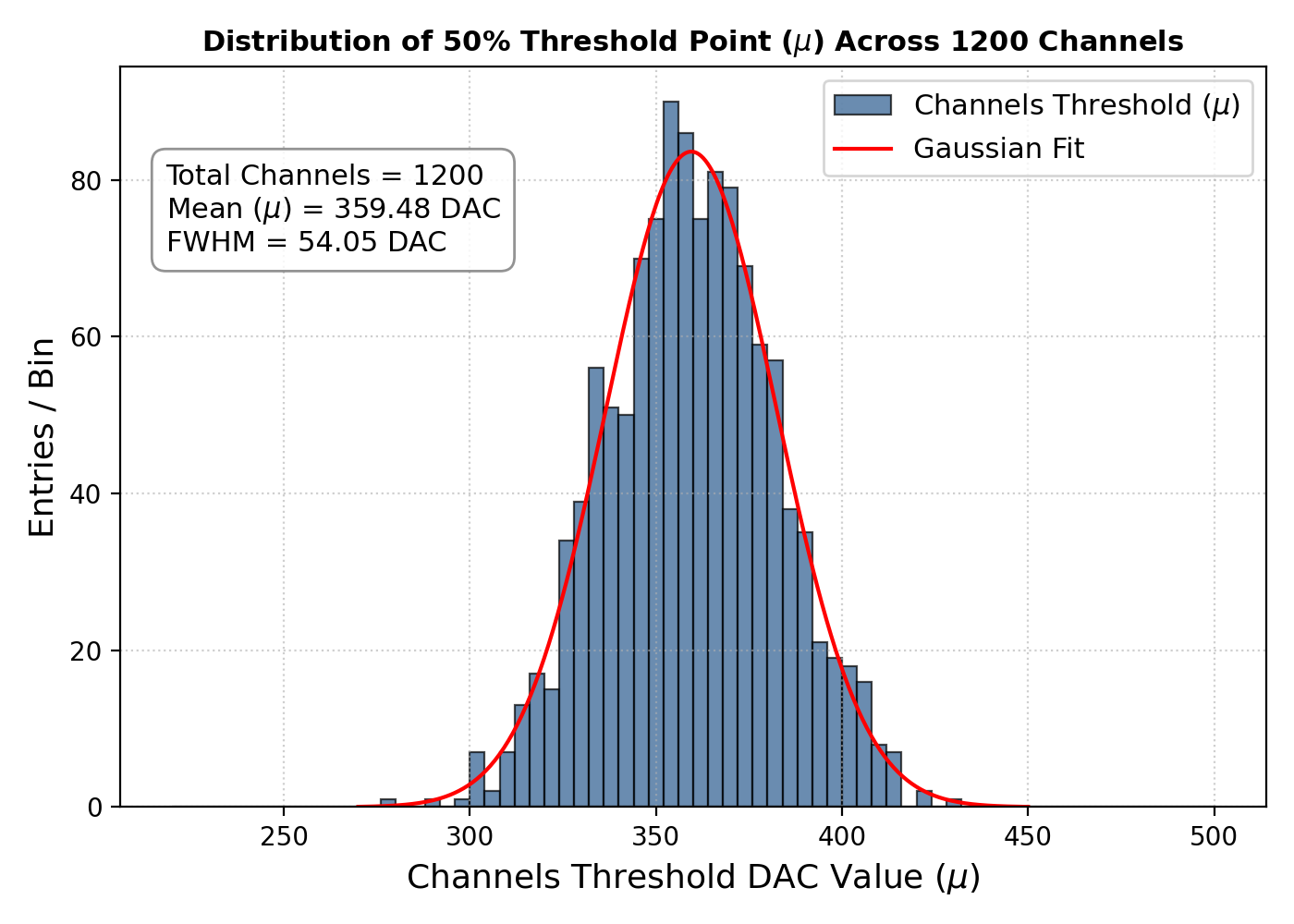}
   	    \label{fig:mu_dis_before_cali}
    }
    \subfigure[]{
  	    \includegraphics[width=.65\textwidth]{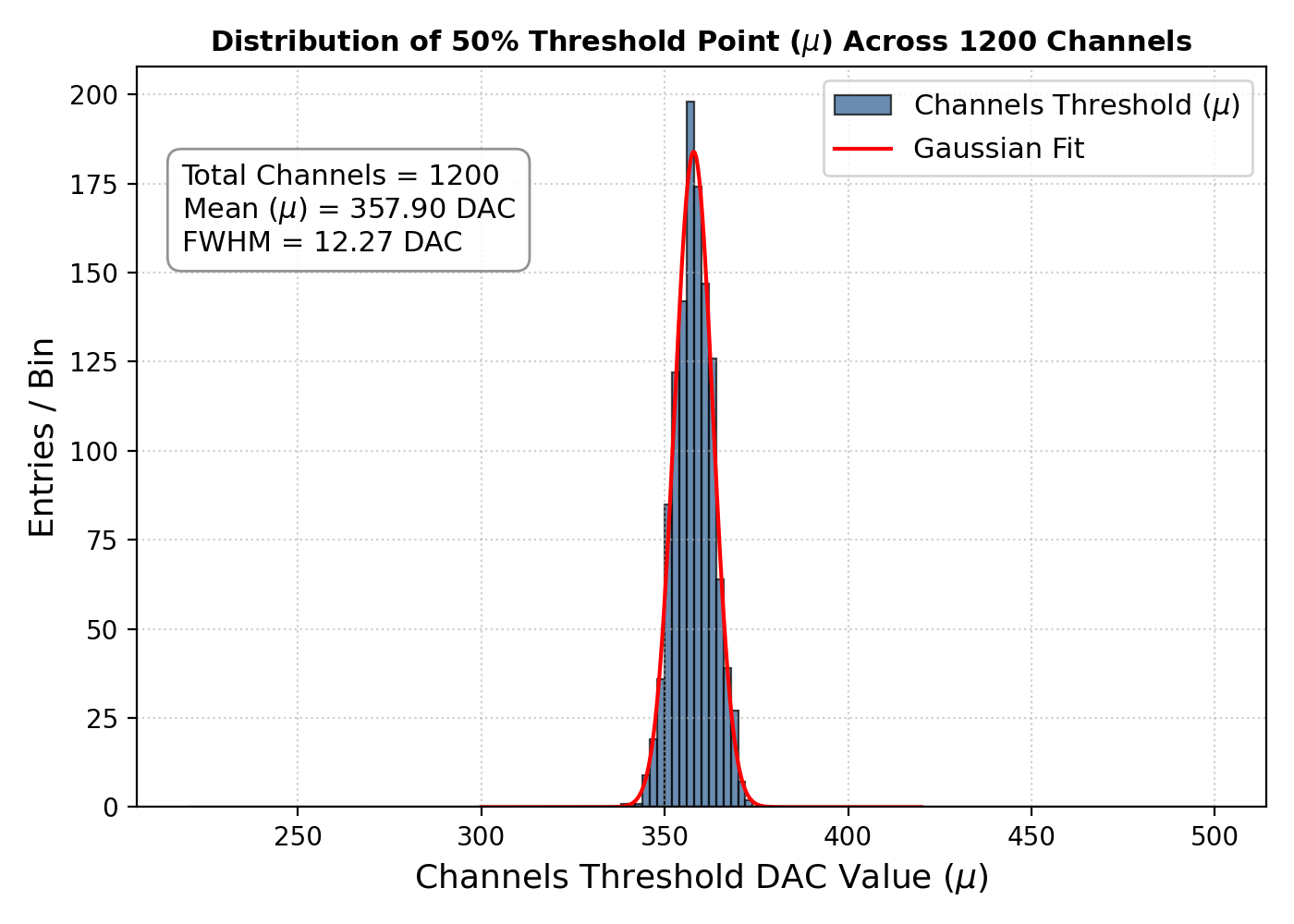}
   	    \label{fig:mu_dis_after_cali}
    }
    \caption{\label{fig:mu_dis}The baseline voltage $\mu$ distribution across 1200 PETIROC channels before (a) and after (b) trim-DAC equalization. The full width at half maximum (FWHM) is reduced from $\sim 54$ DAC units to $\sim 12$ DAC units.}
\end{figure}

The noise rate as a function of threshold voltage $V_{\text{th}}$ follows an Error Function (ERF) distribution, which is the cumulative integral of Gaussian thermal noise:
\begin{equation}
f(V_{\text{th}}) = \frac{N_0}{2} \cdot \left[ 1 - \text{erf}\left( \frac{V_{\text{th}} - \mu}{\sqrt{2}\sigma} \right) \right]
\label{eq:erf_fit}
\end{equation}
where $N_0$ is the saturation hit count, $\mu$ is the baseline voltage $V_{50\%}$ (where the noise trigger efficiency reaches $50\%$), and $\sigma$ represents the equivalent noise charge in voltage units. The host software fits this function online for all channels of one FEB. Before calibration, the baseline values $\mu$ showed a wide spread across the different channels of all PETIROC2Bs, as shown in Figure~\ref{fig:s_curve_before_cali}, which required a higher global threshold to avoid overloading the DAQ with noisy channels. Figure~\ref{fig:mu_dis_before_cali} shows the $\mu$ distribution across 1200 channels before trim-DAC equalization, with a full width at half maximum (FWHM) of $\sim 54$ DAC units ($\sim 50$ mV).

\begin{figure}[htbp]
\centering
\includegraphics[width=0.99\textwidth]{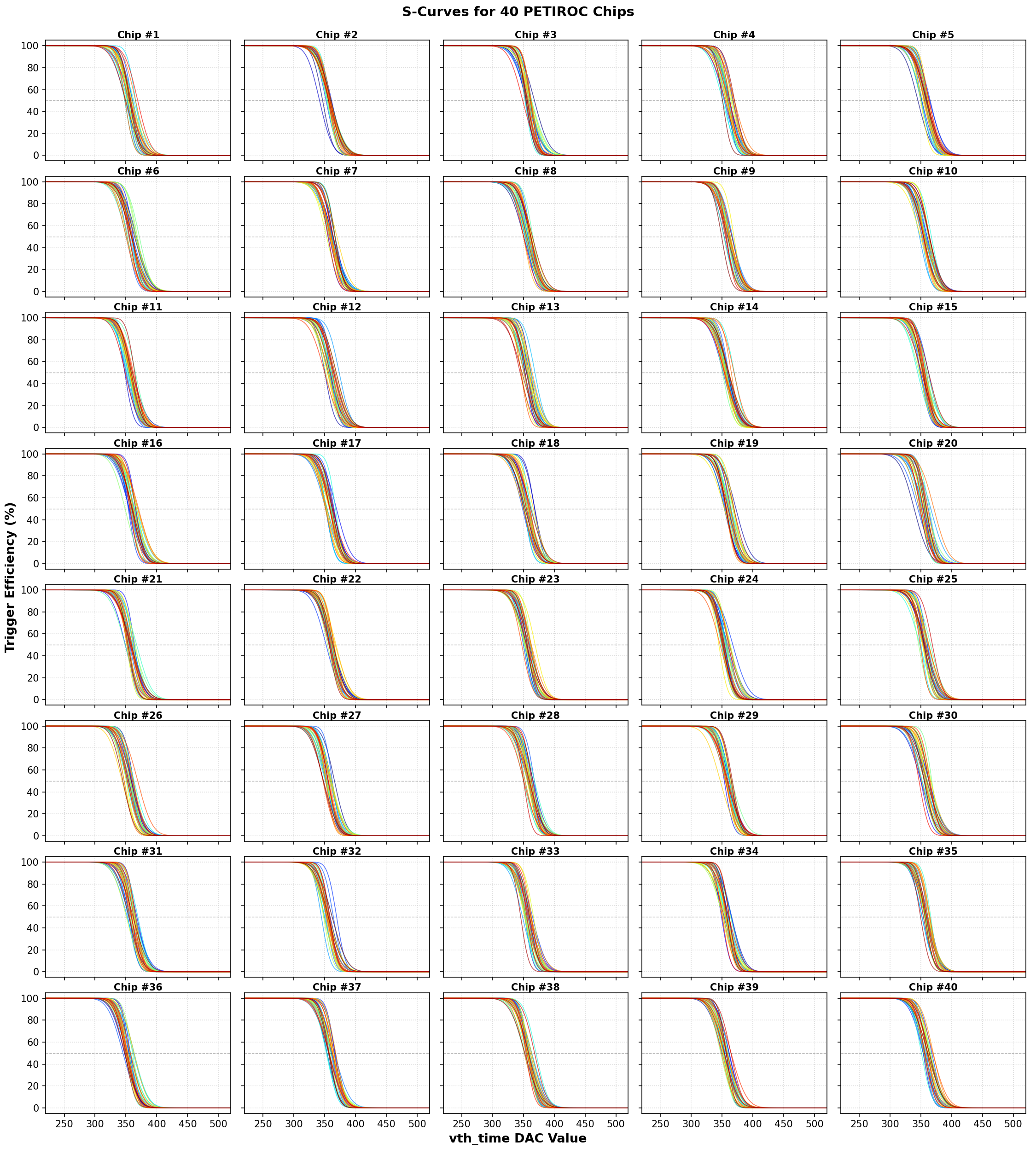}
\caption{The trigger efficiency of 40 PETIROCs as a function of threshold voltage $V_{\text{th}}$ after trim-DAC equalization.}
\label{fig:s_curve_after_cali}
\end{figure}

To flatten the baseline variations, each PETIROC2B channel includes an internal 6-bit fine-tuning DAC (Trim-DAC). Using the fitted $\mu_i$ for each channel $i$, the software calculates the optimal Trim-DAC offset to shift all baselines toward a uniform target voltage $V_{\text{target}}$. After loading these settings into the ASICs, a second S-curve scan confirmed that the baseline spread was significantly reduced, as shown in Figure~\ref{fig:s_curve_after_cali}. The FWHM of the $\mu$ distribution is reduced to $\sim 12$ DAC units ($\sim 11$ mV), as shown in Figure~\ref{fig:mu_dis_after_cali}. This tight alignment allows the system to operate reliably at low thresholds, improving detector sensitivity for minimum ionizing particles.

\subsection{Electronics Timing Performance}
\label{sec:timing_performance}

To evaluate the electronic timing capability and system-level clock distribution precision without detector noise, charge-injection bench tests were performed using a low-jitter programmable pulse generator. A fast-edge pulse signal was split and simultaneously injected into two readout channels to measure the time difference distribution $\Delta t$. Because both channels introduce independent and identical timing jitters, the single-channel electronic time resolution $\sigma_{\text{elec}}$ is extracted as:
\begin{equation}
\sigma_{\text{elec}} = \frac{\sigma(\Delta t)}{\sqrt{2}}
\label{eq:sigma_deduction}
\end{equation}
where $\sigma(\Delta t)$ is the standard deviation obtained from a Gaussian fit to the measured time difference histogram.

\begin{figure}[htbp]
\centering
\includegraphics[width=0.99\textwidth]{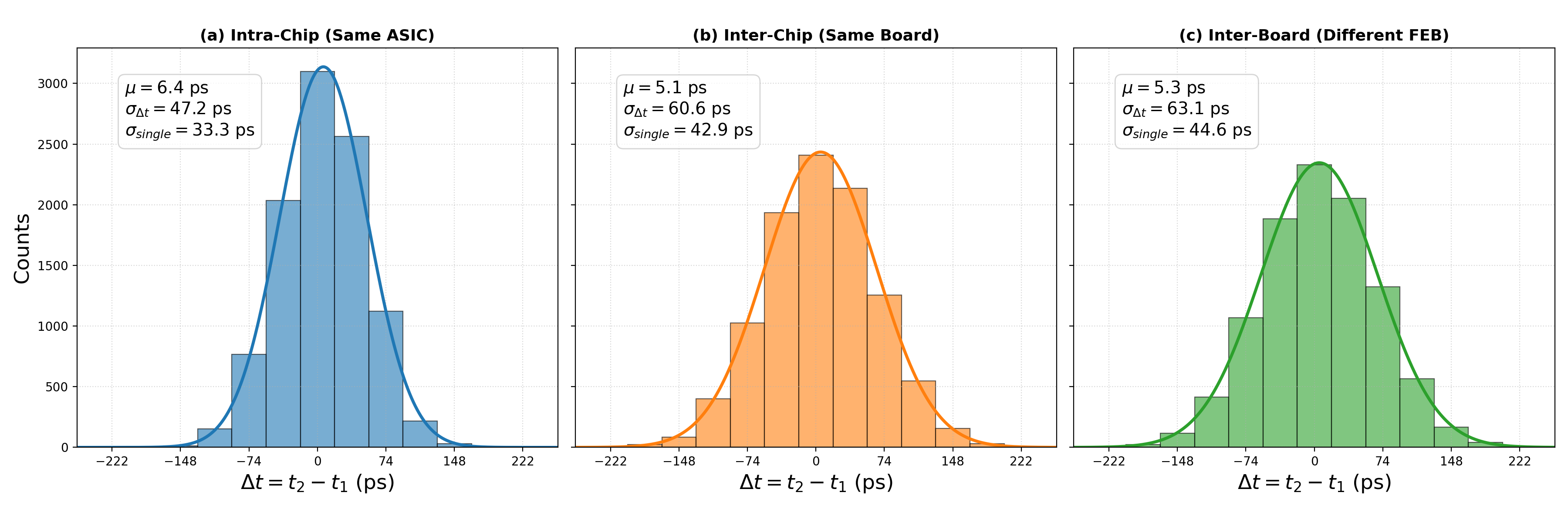}
\caption{Time difference ($\Delta t$) distributions and Gaussian fits for three test configurations: (a) intra-chip, (b) inter-chip (same FEB), and (c) inter-board.}
\label{fig:time_resolution}
\end{figure}

The single-channel timing jitter was first measured by injecting signals into two channels within the same PETIROC2B ASIC. This configuration isolates the intrinsic performance of the front-end preamplifier and discriminator, excluding external clock distribution variations. The measured time difference distribution yielded an intrinsic single-channel electronic time resolution of $\sigma_{\text{intra-chip}} \approx 33\text{~ps}$ RMS, as shown in Figure~\ref{fig:time_resolution} (a), demonstrating the excellent timing performance of the PETIROC2B front end.

To evaluate system-level clock alignment across the system, signals were injected into two chips located on the same FEB and across different FEBs. As shown in Figure~\ref{fig:time_resolution} (b) and Figure~\ref{fig:time_resolution} (c), the extracted inter-chip time resolutions were measured to be $\sigma_{\text{inter-chip}} \approx 43\text{~ps}$ RMS and $\sigma_{\text{inter-board}} \approx 45\text{~ps}$ RMS, respectively. The small degradation between intra-chip and inter-chip timing performance ($\Delta \sigma \approx 10\text{~ps}$) confirms the low clock skew and high phase stability of the clock distribution network.

\section{Conclusions}
\label{sec:conclusions}

In this work, a high-granularity, high-precision timing readout electronics system is developed and validated for a $100 \times 100\text{~cm}^2$ MRPC detector with 2400 readout pads. To achieve high signal integrity, the system adopts an active PCB design that integrates 80 PETIROC2B ASICs directly behind the sensing pads on two modular FEBs. The architecture is supported by FPGA-based data-acquisition firmware featuring multi-channel over-sampling and framing, alongside a flexible distributed slow-control topology that ensures reliable dynamic configuration without compromising physics data bandwidth.

The performance of the system is comprehensively verified through bench tests. The automatic trim-DAC equalization narrows the baseline noise dispersion from 54 to 12~DAC units on 1200 channels of FEB, guaranteeing uniform threshold operation over the active area. Furthermore, signal-injection measurements demonstrate excellent electronic time resolution, achieving $33\text{~ps}$ RMS for single-channel (intra-chip), $43\text{~ps}$ RMS across chips, and $45\text{~ps}$ RMS across boards. These results fully validate the active FEB topology and firmware architecture, establishing a high-performance readout solution for large-area MRPC detectors. Future work will focus on integrating and evaluating these large-scale detectors with the newly developed readout system within the SDHCAL prototype during beam tests.


\acknowledgments

This work would like to thank WEEROC company and Stephane Callier from OMEGA Laboratory for the support on the usage of Petiroc2B ASICs. This work was supported by National Key R\&D Program of China (Grant No. 2023YFA1606203), National Natural Science Foundation of China (Grant No. 12575197), Shanghai Pilot Program for Basic Research — Shanghai Jiao Tong University (Grant No. 21TQ1400218).

\bibliographystyle{JHEP}
\bibliography{myref}
\end{document}